\documentclass{pasa}%

\title[Globular clusters in the Galactic bulge]{Globular clusters
in the Galactic bulge}
\author[Bica et al.]{E. Bica$^1$, S. Ortolani$^{2,3}$, and B. Barbuy$^4$\thanks{This is an example of author footnote}\\
\affil{$^1$Universidade Federal do Rio Grande do Sul, Departamento de Astronomia,
CP 15051, Porto Alegre 91501-970, Brazil}%
\affil{$^2$Dipartimento di Fisica e Astronomia, Universit\`a di Padova, I-35122 Padova,
 Italy}%
\affil{$^3$INAF-Osservatorio Astronomico di Padova, Vicolo dell'Osservatorio 5,
I-35122 Padova, Italy}%
\affil{$^4$Universidade de S\~ao Paulo, IAG, Rua do Mat\~ao 1226,
Cidade Universit\'aria, S\~ao Paulo 05508-900, Brazil}}%
\jid{PASA}
\doi{10.1017/pas.\the\year.xxx}
\jyear{\the\year}

\begin{document}%
\begin{abstract}
A view of the Galactic bulge by means of their globular clusters is 
 necessary for a deep understanding of its formation and evolution.
Connections between the globular cluster and field star
 properties in terms of kinematics, orbits, chemical abundances 
and ages should shed light on different stellar population components.
Based on spatial distribution and metallicity, we define a probable best list
of bulge clusters, containing 43 entries. Future work on newly discovered
objects, mostly from the VVV survey, is suggested. These candidates might
alleviate the issue of missing clusters on the far side of the bulge.
We discuss the reddening law affecting the cluster distances towards
the center of the Galaxy, and conclude that  the most suitable
total-to-selective absorption value appears to be R$_{\rm V}$=3.2, in agreement
with recent analyses. An update of elemental abundances for bulge clusters
is provided.
\end{abstract}
\maketitle%
\section{Introduction }
\label{sec:intro}

The central parts of our Galaxy were prospected by Baade (1946),
in order to detect its nucleus, and to have an indication of
the morphological type of our Galaxy. The 
so-called Baade's Window was revealed, and from variable stars
 identified in the field,
 the bulge stellar population was identified to be similar to 
that defined in Baade (1944) as population II.
In the early 60s the notion of a Galactic bulge was already
established (e.g. Courtes \& Cruvellier 1960).
McClure (1969) found that the bulge stars were super metal-rich compared
 to the stars nearby the Sun.
Whitford and Rich (1983) confirmed the high metal content of bulge stars
 from individual star spectroscopy. 

Catalogues prepared along the decades show a steady growth of
overall samples of globular clusters (hereafter referred to as GCs), 
together with photometric and 
spectroscopic
information. Cannon (1929) used the Harvard plate spectra to
give integrated spectral types of 45 globular clusters.
Mayall (1946) measured radial velocities of 50 globular clusters,
and the integrated spectral types of 40 of them.
 Kinman (1959) listed 32 GCs with indication of
metallicity from their integrated spectra; Morgan (1959) directed efforts
on a relatively bright sample of 13 bulge GCs, now known to be metal-rich.

Surveys with Schmidt plates since the 60s provided the 
Palomar and ESO star clusters, among others.
Terzan (1968 and references therein) significantly increased the number of 
central GCs by reporting new faint ones
in the bulge direction. These discoveries provided 
an important sample of GCs in the bulge.

Other studies of GC overall samples were
presented by van den Bergh (1967), Zinn (1985 and references therein), Bica \& Pastoriza (1983). 
Webbink (1985) provided a catalogue of
 154 GCs and candidates, with
 an important impact on subsequent observational efforts.
 More recently, Harris (1996, updated in 2010,
hereafter Harris10)\footnote{www.physics.mcmaster.ca/~harris/mwgc.dat}
reports properties for 157 GCs.

It is interesting to see how the notion of GCs pertaining to
the bulge and their spatial distribution evolved in the last decades. 
Frenk \& White (1982) found evidence that metal-rich GCs
formed a bar-like structure.
Based on metallicity, scale height, and rotational velocities available at that time, Armandroff (1989 and references therein) 
interpreted a sample of low Galactic latitude metal-rich GCs as belonging to a disk system.
Minniti (1995) instead, from metallicity and kinematics of GCs
in the central 3 kpc (about 20$^{\circ}$) from the Galactic center, found evidence for these GCs to belong to the bulge.
C\^ot\'e (1999) derived metallicities
and radial velocities from high resolution spectroscopy 
for a significant sample of GCs within 4 kpc from the Galactic center. He concluded that metal-rich GCs 
are associated with the bulge/bar rather than the thick disk. 

An important step in the understanding of the nature of the bulge GCs was presented by Ortolani et al. (1995), where
 the metal-rich GCs NGC 6528 and NGC 6553 were found to have an old age, identical to the bulk of the bulge field stars 
and comparable to that of the halo clusters. 
Barbuy et al. (1998) derived new results and summarized the properties
 of 17 GCs in the bulge projected within 5$^{\circ}$ 
of the Galactic center. They concluded that these clusters shared comparable properties with the bulge field stars, 
including not only metal-rich GCs but also intermediate metallicity ones.
 They also found that there are no clusters in
 a strip 2.8$^{\circ}$ wide centred at about 0.5$^{\circ}$ south of
 the Galactic plane.

Several comprehensive recent reviews have addressed the stellar populations, both field and GCs,
in the Galactic bulge (e.g. Harris 2001; Rich 2013; Gonzalez \& Gadotti 2015). These reviews address also comparisons of 
the Galactic bulge with external galaxies.

In the following we describe recent advances on detailed chemical abundances, distances, 
kinematical properties and hints on possible association with subsystems in the central region of the Galaxy.
We also prepare for the future by defining a bulge GC sample,
 which includes suggestions for future studies, in terms
of unstudied objects. In particular it is clear that proper motion derivation
 is still needed for most GCs.

\section{The sample of bulge globular clusters}

In the past, angular distances from the Galactic center were
the basis for selecting bulge GCs in the central parts of the
Galaxy (e.g. Zinn 1985; Barbuy et al. 1998).

As a first approach we selected
clusters with angular distances below  20$^{\circ}$.
Our next step was to use the Galactocentric distances of the clusters, 
given that now they are more reliable - see
Harris10.
 We adopted a distance to the Galactic center of 8 kpc
 (Bobylev et al. 2014; Reid et al. 2014).
For selecting the clusters we have experimented different distances from the Galactic center
 of 6, 5, 4 and 3 kpc. In each case, we checked for bulge clusters and halo intruders.
We conclude that a cutoff of 3 kpc is best in terms of isolating a bona-fide
bulge cluster sample, with little contamination.
We finally applied a metallicity filter: Zoccali et al. (2008), Hill et al.
(2011), Ness et al. (2013), Rojas-Arriagada et al. (2014)
 have shown that the lower metallicity end of the
bulge is around [Fe/H]=-1.3.
This is confirmed with the findings by Pietrukowicz et al. 2012, 2015),
D\'ek\'any et al. (2013) and Lee
 (1992), all of them having demonstrated that there is a peak of RR Lyrae
with [Fe/H]=-1.0, centrally concentrated and spheroidal
(except for Pietrukowicz et al. 2015 that found it to be elongated), 
corresponding to an old bulge. We verified 
 that GCs with [Fe/H] $<$ -1.5 corresponded to well-known 
halo clusters in most cases.
 
Criteria using the space velocity V, as proposed by Dinescu et al.
(1997, 1999a,b, 2003),
and Casetti-Dinescu et al. (2007, 2010, 2013), are only feasible 
if proper motions are available,
 besides radial velocities. In particular Dinescu et al. (2003)
verified the classification of clusters
as members of different galaxy components in terms of kinematics.
The use of this criterion is possible  for about a third of the GCs, 
 at the present stage. GAIA results in a few years
will bring higher precision, and new data for a number of GCs.
A few groups are applying proper motion cleaning to bulge clusters
(e.g. Dinescu et al. 2003; Rossi et al. 2015a).
Therefore Carretta et al. (2010) criterion
of space velocity V values to select stellar populations,
is not possible presently for having only the radial velocity 
in many cases, but may be applicable in a few years.

In Table \ref{tab1} we present the selected bulge clusters, 
following the criteria
explained above. Djorgovski 1 is added to the list, because it was in the
Barbuy et al. (1998)'s list, and the distance available in Harris10, based
on infrared data, is probably overestimated. The clusters are ordered
by right ascension, to be compatible with Harris10.

The distances from the Sun are from  Harris10 mainly based
on the visual magnitude of the HB, while Valenti et al. (2007) give distances
based on isochrone fitting in JHK.
 We also report distances derived from the Galactic center, as given by
Harris10. For Kronberger 49, not listed in Harris10, data are from
Ortolani et al. (2012).
 The distance derivation is crucial in order
to get the position of the clusters relative to the Galactic center and the bar.
In most cases, for a cluster located inside the bar, its orbit is trapped,
 given the high mass of the bar. In fact more than half the bulge
mass is in the bar, and for this reason, the bulge clusters that show
low kinematics are trapped (Rossi et al. 2015a). 
Given the old age of the clusters
it is likely that they were formed before the bar, and later trapped.
The dynamical behaviour of such clusters was illustrated for HP~1
 (Ortolani et al.
2011), and for 9 GCs by Rossi et al. (2015b),
 all of them showing a low maximum height. 

\subsection{Intruders and missed bulge clusters?}

In Table \ref{tab2} we list possible intruders to our main selection, as well
as limiting cases, and  we collected a few more GCs with other
 distance and metallicity criteria. Subsamples are classified as follows:

a) Probable halo intruders with [Fe/H]$<$-1.5: 
we found 6 GCs with distances
to the Galactic center smaller than 3 kpc (our selected bulge volume), 
but with metallicity lower than [Fe/H]$<$-1.5, and besides with very high spatial velocities
(Dinescu et al. 1999a, 2010), all pointing to perigalactic locations of halo GCs.

b) Outer bulge shell with distances 3$<$R$<$4.5 kpc: this surrounding bulge shell
 does probably contain
true bulge GCs, that should be near apogalacticon. Interestingly, the few
space velocities availabe (Dinescu et al. 1999a, Cudworth \& Hanson 1993) 
for this sample are rather low.

c) Outer bulge  shell intruders: 
five GCs are located in this shell b), but showing low
metallicities, as for example M22, having also a high velocity typical of halo, 
like in group a).

d)  Metal-rich GCs ([Fe/H]$>$-1.0) beyond R$>$4.5kpc: this sample corresponds to
 the ones that conveyed the idea of ``disk globular clusters''
(e.g. Armandroff 1989). Many of these key and well-known GCs do not have
space velocities derived so far. This would be of great interest to constrain
the possibility to link them to the thick disk.

e) Intruders to d). Data for FSR 1767 are from Bonatto et al. (2007, 2009).

f) Little studied GCs without enough parameters: probably key objects
for future analysis (Sect. \ref{sect7})

\subsection{Multiple population clusters in the bulge\label{Sect22}}

Terzan 5 was identified to have at least two stellar populations
and given its high mass, it was proposed to be a stripped dwarf galaxy
(Ferraro et al. 2009; Massari et al. 2014). 
Saracino et al. (2015) concluded that Liller 1 is
as massive as Terzan 5 or $\omega$ Centauri. These massive clusters have
absolute magnitudes around M$_{\rm V}$$\approx$-10. 

On the other hand,
the faintest bulge GCs are as faint as  M$_{\rm V}$$\approx$-4, such
as Terzan 9 and AL~3. These magnitudes are comparable to those of the ultra-faint
galaxies (e.g. McConnachie 2012; Bechtol et al. 2015).

By multiple stellar populations we refer to different metallicities
[Fe/H] in a same cluster. For most clusters there are hints of
two populations, but no difference in metallicity.
A comprehensive discussion on Na-O anticorrelation was given
by Gratton et al. (2015), where differences between red and blue horizontal
branch stars allow to derive some important conclusions. 
See further discussions on multiple populations revealed by Na-O 
anticorrelations in Sect. 4.  Since this effect appears to be
 present for most clusters, this is incorporated in the 
definition of globular clusters, and it is not a concern in the present work.

\section{Distances\label{sect3}}

Distances of bulge GCs are mainly based on the horizontal branch 
(HB) luminosity level.
 The calculation of distances depends on 3 basic inputs: 
(1) the HB absolute magnitude,
 (2) the reddening, and (3) the reddening law. These three datasets 
slightly depend on the metallicity. 
The HB absolute magnitude may also depend on the He abundance. 
The current statistics is based mostly on the distances 
reported in Harris10, where the HB level is given by
the relation: M(HB)=0.16[Fe/H]+0.84. In Harris10 if the HB level
is not available, specific references are used.
This relation is somewhat different from that used by
Barbuy et al. (1998) in their catalogue
of 17 inner bulge clusters, adopted from Jones et al. (1992):
M(HB)=0.16[Fe/H]+0.98. This latter relation gives an
HB level of about 0.14 magnitudes fainter, producing as
well smaller distance moduli, but this difference is
negligible when compared to other uncertainties.
It is also important to point out that the main difference
between Harris10 and Barbuy et al. (1998) values, is that
the latter was based on optical CMDs only.
The list of distances also makes use of the JHK Colour-Magnitude Diagrams 
(CMDs) when it is considered more reliable.
Finally the catalogue by Bica et al. (2006) uses basically the same 
assumptions as Harris10.

{\it Reddening and reddening law:}
The reddening is a key parameter in the derivation of the distances based on the photometric technique. 
For most of the low latitude bulge GCs, an error of 5\%  in reddening produces a typical error of
 $\sim$15\% in the absorption, which propagates with the same fraction to the distance modulus.
In Harris10 the reddening has been derived from 
Webbink (1985), Zinn (1985) and Reid et al. (1988).
 The standard reddening law (R$_{\rm V}$=3.1) 
has been used to derive the visual absorption.
Bica et al. (2006) uses basically the same input values, 
but they converted E(B-V) into A$_{\rm V}$ using
 R$_{\rm V}$=3.1 for clusters with [Fe/H]$<$-1.0 and R$_{\rm V}$=3.6 for 
[Fe/H]$>$-1.0 following Grebel and Roberts (1995), and 
R$_{\rm V}$ values have been interpolated in the intermediate 
metallicity interval. 
In principle this choice should produce shorter distances than Harris10 for high metallicity clusters.
A different computation was performed for most of the inner bulge clusters presented in Barbuy et al. (1998).  
Eq. A1 of Dean et al. (1978) was used to convert E(V-I) to E(B-V) and then the reddening
 dependence of R$_{\rm V}$ on E(B-V) as given in Olson (1975) has been used to
 convert E(B-V) into A$_{\rm V}$: R$_{\rm V}$=3.1+0.05([Fe/H]).

Independent distances and reddening values
 have been obtained by Valenti et al. (2007) for 37 bulge clusters,
 using infrared JHK photometry. The main advantage of the JHK derived distances is that the reddening vs.
 absorption ratio is almost constant even if R$_{\rm V}$ varies (Fitzpatrick 1999; Cardelli et al. 1989).
The comparison of the optical vs. IR distances give 
a chance to probe the reddening law,
 in the optical regime, in the direction of the Galactic bulge. 

Figure \ref{Fig2} shows the distance differences between different
R$_{\rm V}$ values versus reddening.
The absolute magnitude of the horizontal branch M(HB)
is assumed to be of M(HB)=0.68 for [Fe/H]=-1.0,
at E(B-V)=0.
Therefore at the distance of the Galactic center,
assumed here to be of 8 kpc, the distance modulus is
m-M=14.5, and m(HB)=14.5+0.68=15.18.
So for the zero point we assume: E(B-V)=0, m(HB)=15.18.
We use 12 clusters in common with Valenti et al. (2007),
and adding Terzan 9 (Ortolani et al. 1999). From this sample, 
 an average distance difference of  d(IR)-d(optical) =  0.65 kpc has been derived. 
In the infrared sample we have also 3 clusters (Terzan 4, Terzan 5 and NGC 6528) with data from HST/NICMOS
 (Ortolani et al. 2007). In these cases we adopted an average value between NICMOS and Valenti et al. (2007).
At an average distance of 8 kpc,
 the 0.65 kpc difference is equivalent to about 0.3 mag. in distance modulus.
Conclusions from Fig. 2 are: a)
it is clear that the optical data produce shorter distances,
in particular at high reddening; b) in order to get
comparable infrared and optical distances, we
have to adopt an average total-to-selective absorption R$_{\rm V}$=3.2. 
This average value is slightly lower than that adopted in Barbuy et al. (1998)
 of R$_{\rm V}$=3.39 (see their Table 4), but it is still higher than the standard R$_{\rm V}$=3.1 value, and it is in agreement 
with recent studies of the reddening law in different conditions of reddening, metallicity and intrinsic
 colours of the stars (Hendricks et al. 2012; McCall 2004). 
A further test can be performed following 
Racine \& Harris (1989) and Barbuy et al. (1998), 
plotting  V(HB) as a function
 of reddening. Assuming that, on the average, the inner bulge clusters (within 3 kpc from the Galactic 
center) are concentrated around this distance, the HB level should be related to the reddening 
by means of a slope R$_{\rm V}$. This is plotted in
Fig. \ref{Fig2}, which is the updated version of Fig. 5 in Barbuy et al. (1998). 
This analysis includes 12 clusters plus an arbitrary
 point at E(B-V)=0 and V(HB)=15.18, which should correspond to the V(HB) at 8.0 kpc from the Sun, with [Fe/H]=-1.0,
 M$_{\rm V}$=0.8, as explained above. Therefore the best fit is 3.3 $<$ R$_{\rm V}$ $<$3.1. The higher slope of
 R$_{\rm V}$=3.6 seems too steep. This confirms that a choice of an average reddening
value of R$_{\rm V}$=3.2 is adequate 
and that the baricentric distance from the Sun of the considered clusters should be around 8.0 kpc. 

Finally, we address a few comments on the distance of UKS~1, since there is
disagreement between different authors. We employed 4 sets of data:
a) NICMOS data (Ortolani et al. 2001, 2007), measured relative to NGC 6528,
assumed to be at a distance of d$_{\odot}$ =  7.7 kpc, results in 
(m-M)$_{\circ}$= 14.43. Assuming that both clusters have a similar metallicity,
and the HST/NICMOS reddening law, we get  d$_{\odot}$ = 15.8 kpc for UKS~1;
b) NICMOS data measured relative to Liller 1, assuming for Liller 1
a distance of  d$_{\odot}$ = 8.1 kpc (from (m-M)$_{\circ}$=14.55, Saracino
et al. 2015); given a $\Delta$(m-M)$_{\circ}$=1.02, we get
(m-M)$_{\circ}$=15.57 and a distance of  d$_{\odot}$ = 13.0 for UKS~1;
c) Assuming Minniti et al. (2011) absolute reference for the HB colour
and magnitude M$_{\rm K}$=-1.65, J-K=0.71 (from a study of red clump 
calibrated stars with Hipparcos data), and using NICMOS calibrated data,
we obtain J$_{\rm HB}$=17.96. From M(HB)$_{\rm J}$ = -0.94 and A$_{\rm J}$=2.78,
obtained from the comparison with Liller 1 (Saracino et al. 2015), we
get (m-M)$_{\circ}$=18.63-2.78=15.85, and d$_{\odot}$ = 14.8 kpc; d)
from Liller 1 (Saracino et al. 2015), but with a recalculated absolute
distance from Minniti et al. (2011) values of M(HB)$_{\rm K}$ and J-K(HB),
we have:  (m-M)$_{\circ}$(Liller 1)=14.85(d$_{\odot}$ =9.33 kpc),
and (m-M)$_{\circ}$=14.85$\pm$1.02=15.87 and d$_{\odot}$ = 14.9 kpc for
UKS~1. All these methods give an average of d$_{\odot}$ = 14.6 kpc
for UKS~1, as reported in Table \ref{tab1}.

\section{Chemical abundances}

The metallicity distribution of bulge clusters as given in Table \ref{tab1} is
shown in Fig. \ref{Fig6}. It shows a peak around [Fe/H]$\approx$-1.0, 
suggesting that this population is intrinsically significant. That such
 an old bulge stellar population
is important, is confirmed by studies of RR Lyrae by Lee (1992), D\'ek\'any et al.
(2013), Pietrukowicz et al. (2012, 2015). Also, from the metallicity distribution
function of field stars by Zoccali et al. (2008), Ness et al. (2013),
and Rojas-Arriagada et al. (2014),  this
moderate metallicity corresponds to the lower end of the
metallicity distribution function (MDF) of the bulge bulk
field stellar population. 

Table \ref{tab3} shows the chemical abundances for a subset of GCs from 
Table \ref{tab1}, that have available high spectral
resolution abundance analyses.

{\it Carbon and Nitrogen:} C and N show the expected
anticorrelation due to CNO processing along the red giant branch.
Two clusters with low N abundances should be further studied since this
is not expected.

{\it Odd-Z elements Na, Al}
A crucial issue concerns possible Na-O anticorrelation, which would
indicate the presence of a second stellar generation. As a matter of fact,
most GCs are presently being found to have at least two
stellar generations, except possibly the least massive clusters.
A threshold in mass for a second generation not to occur is presently
estimated to be at 3-4$\times$10$^4$ M$_{\circ}$ (R.G. Gratton, private
communication), or in other words, essentially all GCs should show 
the Na-O anticorrelation.
The second generation is detected via Na-O, Mg-Al anticorrelations,
and also by the presence of both CN-strong and
 CN-weak stars (Carretta et al. 2009).
The origin of these chemical anomalies is probably hot
bottom burning (HBB) (Ventura et al. 2013)
in massive Asymptotic Giant Branch (AGB) first generation stars, with yields ejected in the 
internal cluster medium, and incorporated by
the second generation stars, the latter showing these anomalies. A 
thorough discussion on the origin of multiple stellar populations
in globular clusters is given in Renzini et al. (2015).
From Table \ref{tab3} we see that very few of the bulge clusters
were investigated in terms of Na and Al. In fact, four stars analysed
by Barbuy et al. (2014) show no Na-O anticorrelation (see their
Fig. 6), however more stars have to be analysed for a firm conclusion.
An important investigation on this matter was carried out by Gratton
et al. (2015), where 17 BHB and 30 RHB stars of NGC 6723 were analysed. It was
found that RHB and intermediate-BHB stars appear to belong to a first
generation, showing O-rich and Na-poor abundances, whereas the bluest
of the BHB stars show lower oxygen and higher Na (four BHB stars
show [O/Fe]=+0.23 and [Na/Fe] = +0.11), in contrast with the mean values
given in Table \ref{tab3}. Gratton et al. (2015) consider that extended
blue HB stars might correspond to a second generation of lower mass stars.

{\it Alpha-elements O, Mg, Si, Ca, Ti}
 In Fig. \ref{Fig7} are plotted the abundances of these alpha
elements vs. [Fe/H], including the abundances for the bulge sample, as
given in Table \ref{tab3}, compared with abundances for 57 field bulge
giant stars
from Lecureur et al. (2007), Gonzalez et al. (2011) and Barbuy et al.
(2015), and 58 bulge dwarf stars from Bensby et al. (2013). 
The oxygen abundances are as given in Barbuy et al. (2015), where
they were revised with respect to Lecureur et al. (2007).
Some discrepancy between the oxygen abundances from Barbuy et al. (2015) and
Bensby et al. (2013) can be explained by the facts that:
Barbuy et al. analysed red giants, using the forbidden [OI]6300.31 {\rm \AA} line
and Bensby et al. (2013) analysed dwarfs, using the permitted triplet OI lines
at 7771.94, 7774.16, 7775.39 {\rm \AA} lines. Solar oxygen abundances adopted
were respectively $\epsilon$(O) = 8.87 and 8.85 for Barbuy et al. and Bensby
et al., which would tend to invert the small shift between the two sets results. 
All in all, given that the permitted OI lines are well-known to be subject to
strong non-LTE effects, and tend to overestimate the oxygen abundances, we 
can consider that there is a good agreement between the two sets of oxygen
abundances. 

The panels in Fig \ref{Fig7} indicate therefore that the alpha-elements 
O, Mg, Si and Ca are overabundant by about 
[O,Mg,Si,Ca,Ti/Fe]$\approx$0.3 to 0.4 dex for the more
metal-poor stars. The same is found 
for the field (see also Alves-Brito et al. 2010; Bensby et al. 2013).
This implies early
fast enrichment by core collapse supernovae SNII, which in turn give a short
timescale for bulge formation, 
where even the metal-rich clusters are old (Ortolani et al. 1995).

{\it Heavy elements}
Very few heavy element abundance derivation is available for individual stars
of bulge GCs. The heavy elements of first peak Y, Sr, Zr, together with a few
elements of the second peak Ba, La, and the r-element Eu, can reveal the nature
of the first stars, or else to reveal if AGBs were acting as important 
chemical contributors. A threshold of about [Ba/Eu]=0.6 would be indicative
if Ba was produced by r- or s-process. If [Ba/Eu]$>$0.6, this
 could be a hint of early enrichment by massive spinstars, and
 the same applies to the ratios Y/Ba, Sr/Ba, and Zr/Ba  
(Frischknecht et al. 2015). 
These ratios were studied for the globular cluster NGC 6522,
by Barbuy et al. (2009), Chiappini et al. (2011), and Ness et al. (2014)
using GIRAFFE spectra from the survey by Zoccali et al. (2008).
Barbuy et al. (2014) used new UVES spectra observed in 2012
 for four of the same stars,
and these new results are reported in Table \ref{tab3}.
Sr lines are unreliable as shown by Barbuy et al. (2014),
and Ness et al. (2014), superseding abundance values given
in Chiappini et al. (2011). 
Zr is difficult to derive: in Table \ref{tab3} it is not reported
given that it is only available for NGC 6522, but it is worth
mentioning that Cantelli et al. (2015, in preparation)
 seems to detect a clear variation in Zr
abundances among 12 member stars of NGC 6522 observed with GIRAFFE in 2012.
 The ratios from first to second peak of heavy elements
can also be explained by
 production of s-process heavy elements in massive AGB stars,
and predicted relative ratios
between different heavy elements was presented in Bisterzo et al. (2010, 2013).
Finally, it is possible that all heavy elements in old stars were produced by
the r-process only, as first suggested by Truran (1981).

\begin{figure*}
\begin{center}
\includegraphics[width=17cm]{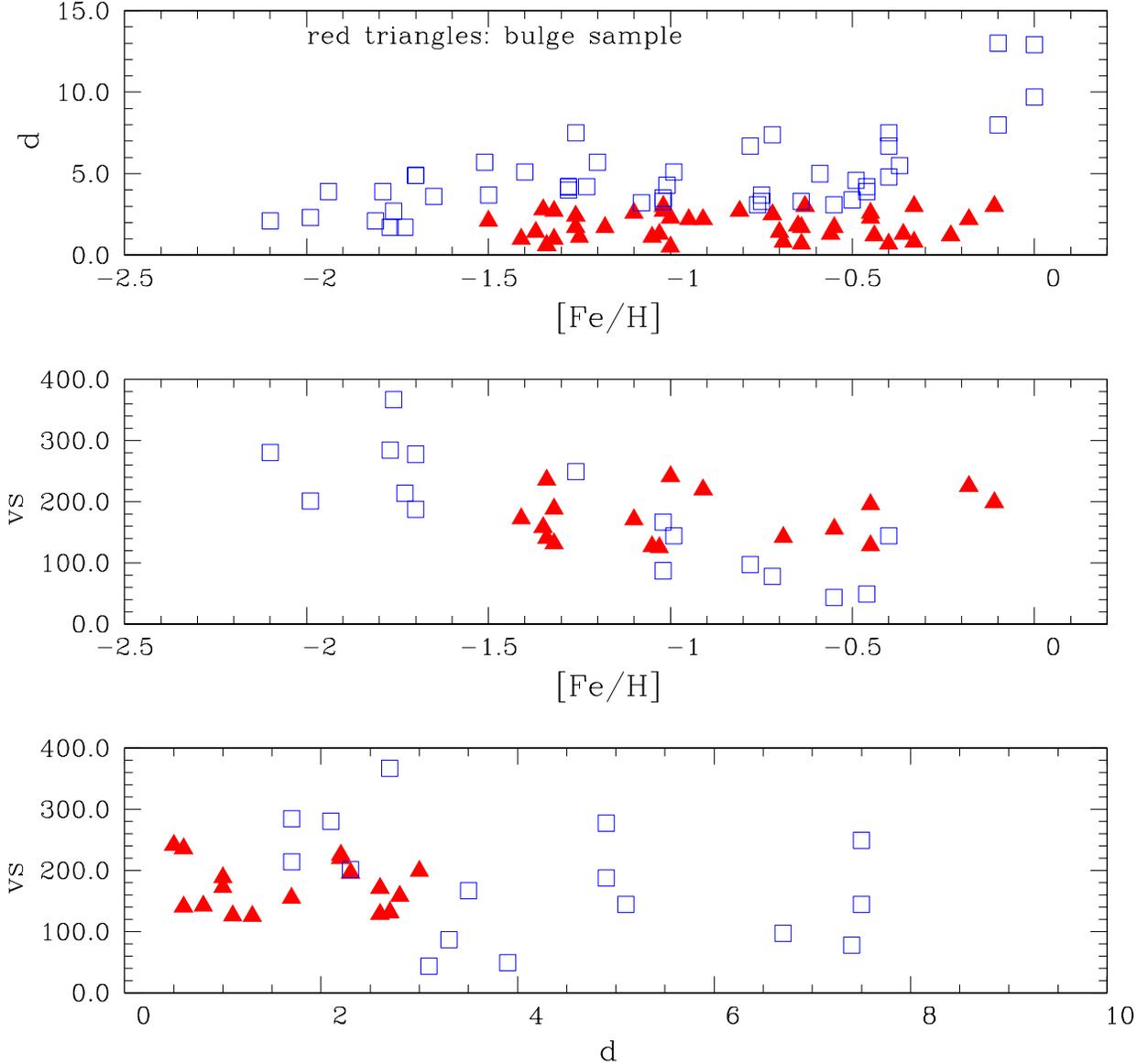}
\caption{Comparisons between Galactocentric distances d(kpc), space velocities V$_{\rm s}$
(km s$^{-1}$) and metallicity [Fe/H]. Symbols: red triangles: bulge clusters (Table \ref{tab1});
open squares: intruders or candidates (Table \ref{tab2})}.
 \label{Fig1}
\end{center}
\end{figure*}

\begin{figure*}
\begin{center}
\includegraphics[width=10cm,angle=0]{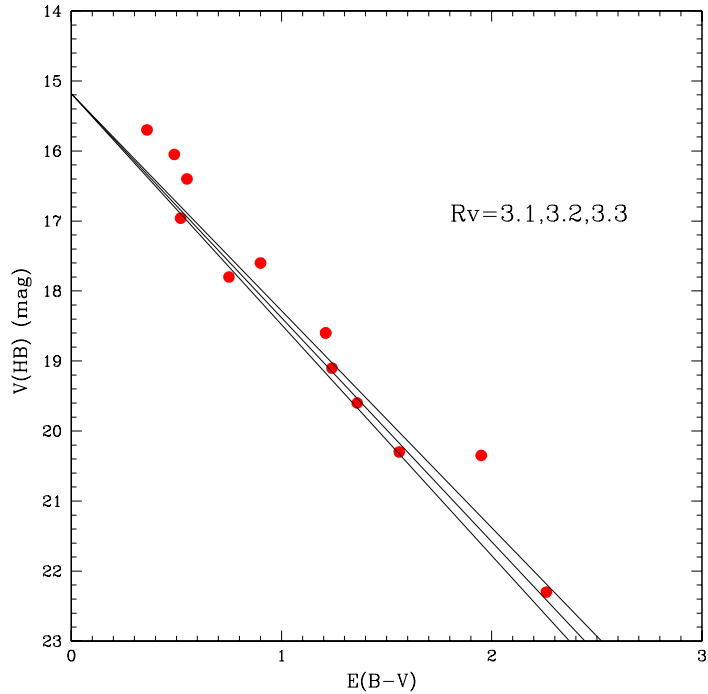}
\caption{Horizontal branch magnitude V(HB) vs.
 reddening E(B-V) for the bulge clusters.
Total-to-selective absorption R$_{\rm V}$  values are indicated.
A distance to the Galactic center of 8 kpc is assumed.}
 \label{Fig2}
\end{center}
\end{figure*}

\begin{figure*}
\begin{center}
\includegraphics[width=10cm,angle=0]{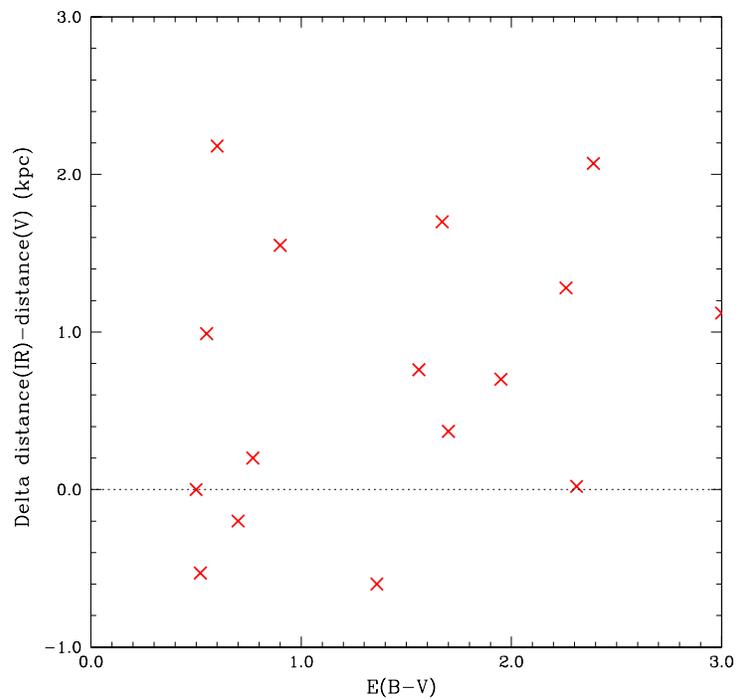}
\caption{Difference of IR vs. optical distances as a function of
reddening for a sample of
bulge clusters in common between Barbuy et al. (1998) and Valenti et al. (2007). }
 \label{Fig3}
\end{center}
\end{figure*}

\begin{figure*}
\begin{center}
\includegraphics[width=10cm,angle=0]{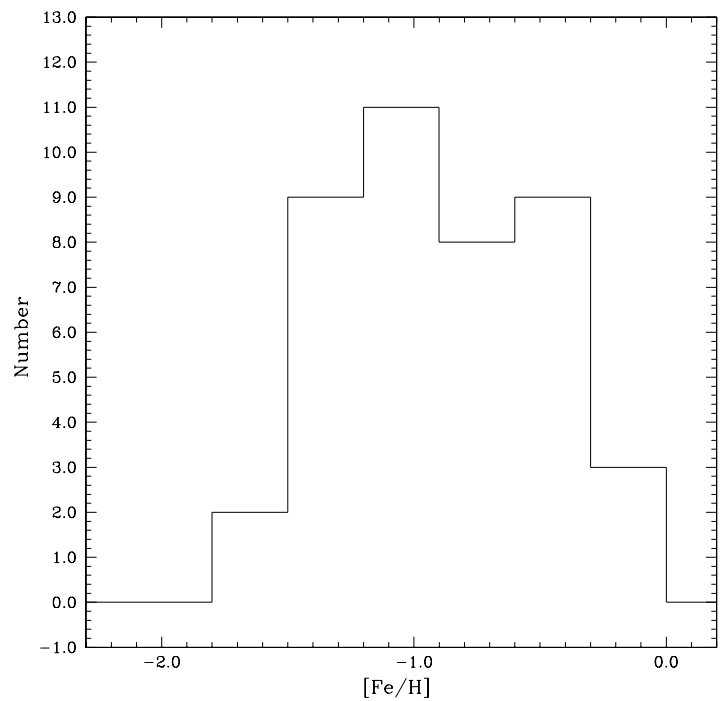}
\caption{Metallicity histogram of sample bulge clusters (Table \ref{tab1}). }
 \label{Fig4}
\end{center}
\end{figure*}

\begin{figure*}
\begin{center}
\includegraphics[width=17cm]{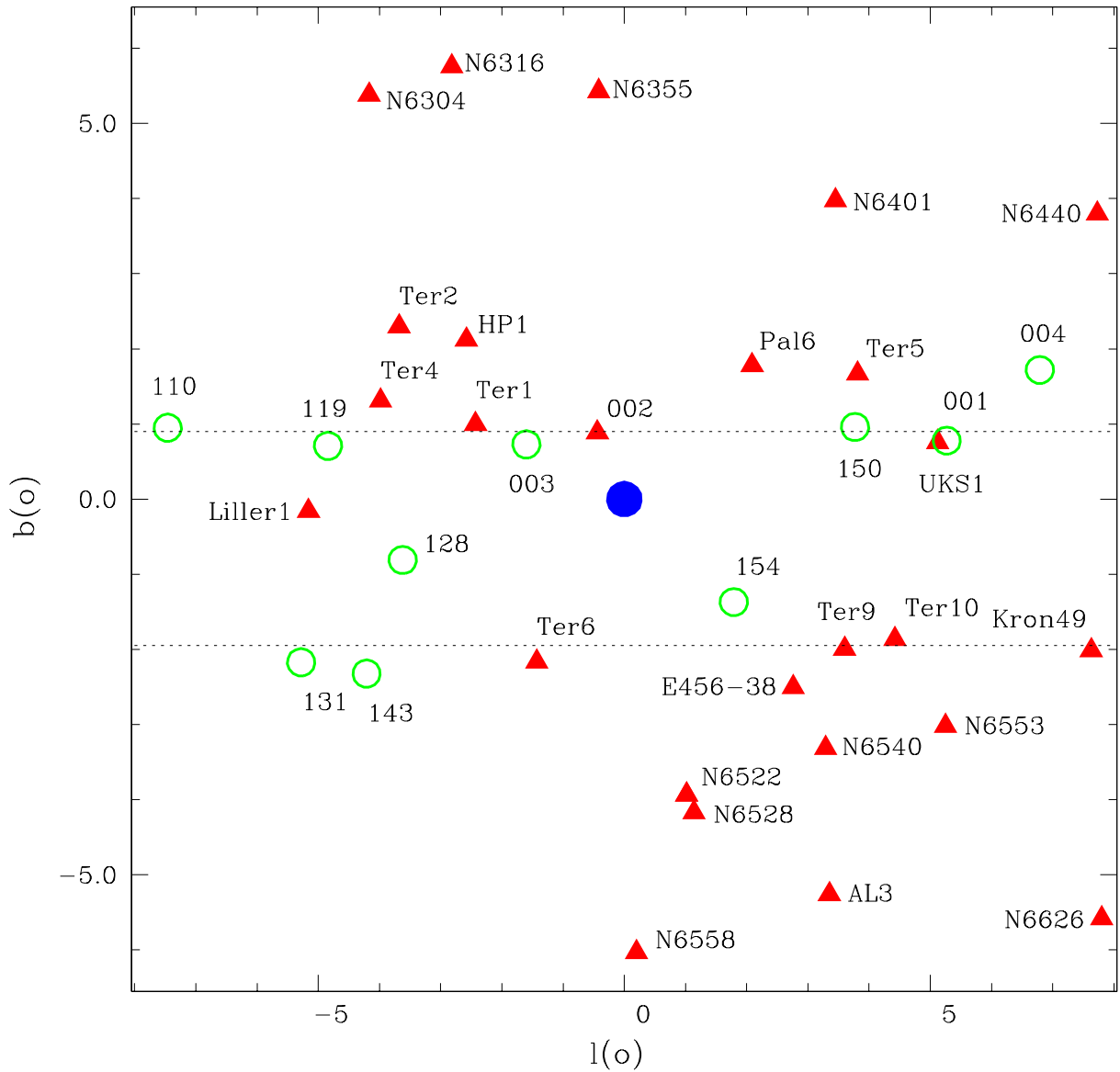}
\caption{Location of central projected bulge clusters in Galactic coordinates.
Symbols: red filled triangles are bulge globular clusters (Table \ref{tab1}); 
green open circles: VVV clusters and candidates (Table \ref{tab4}). VVV clusters
are identified by their numbers; blue filled circle:
Galactic center; dotted lines encompass the so-called forbidden zone for
optical globular clusters.}
 \label{Fig5}
\end{center}
\end{figure*}

\begin{figure*}
\begin{center}
\includegraphics[width=12cm,angle=0]{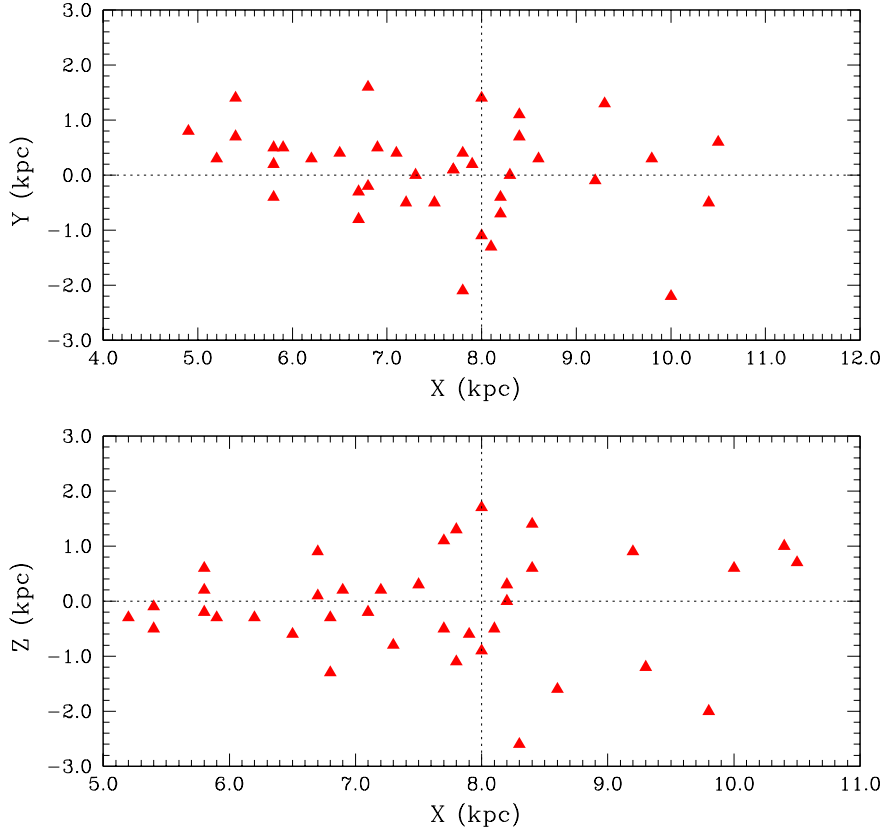}
\caption{Bulge sample (Table \ref{tab1}): Upper panel: in the X,Y plane;
Lower panel: X,Z plane.}
 \label{Fig6}
\end{center}
\end{figure*}

\begin{figure*}
\begin{center}
\includegraphics[width=17cm,angle=0]{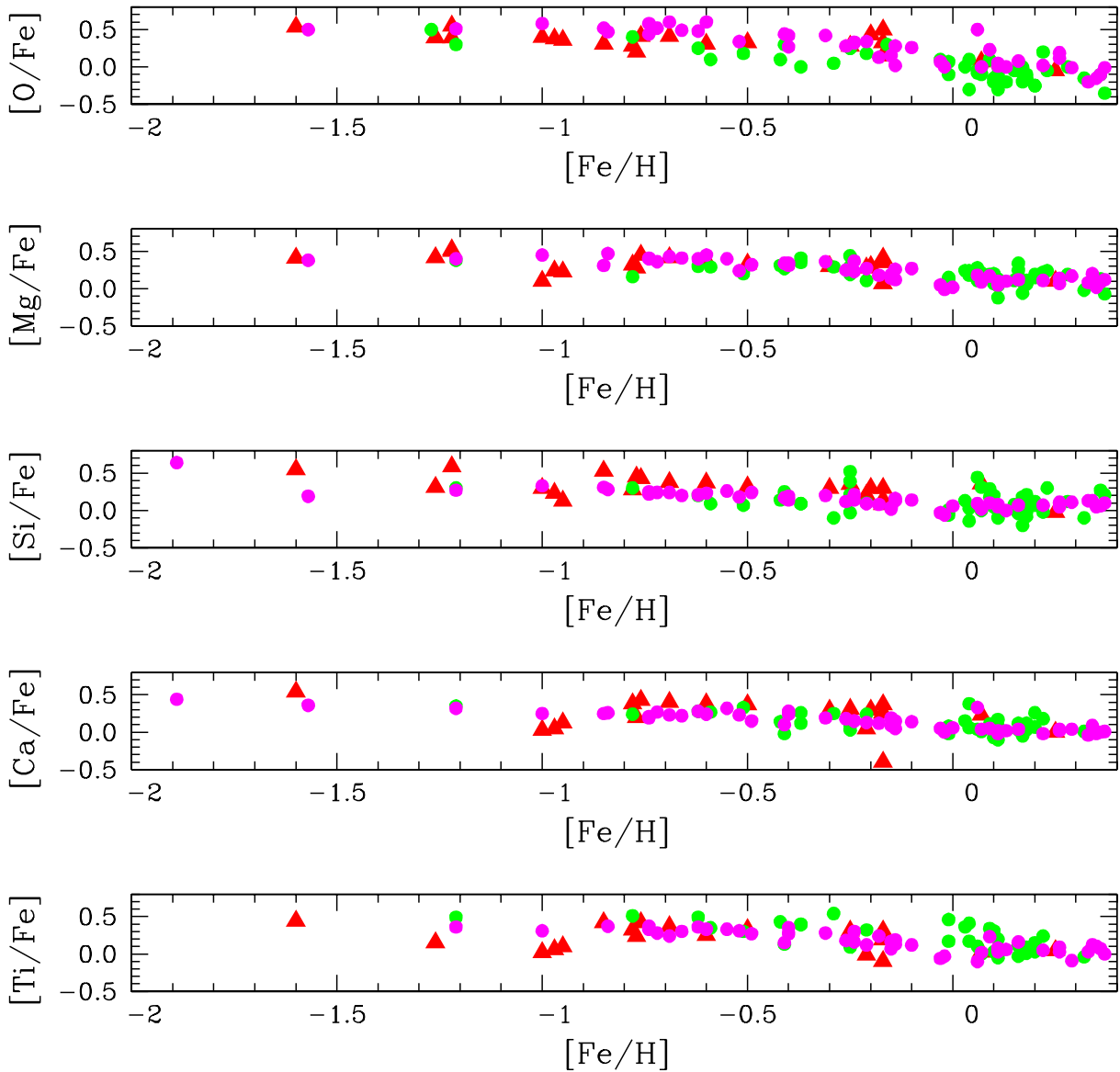}
\caption{[O,Mg,Si,Ca,Ti/Fe] vs. [Fe/H] for sample globular clusters
(red filled triangles), compared with field stars.
For Terzan 5, NGC 6528, NGC 6553, and NGC 6723, 2, 3, 3,
and 2 sets of values are
included, respectively.}
 \label{Fig7}
\end{center}
\end{figure*}

\begin{table*}
\caption{Bulge globular clusters. Reddening values in columns 4, 5 and 6 correspond to Harris10,
Valenti et al. (2007), and our studies along the years. Galactocentric distances and metallicities
are from Harris10, except for UKS~1 (see Sect. \ref{sect3}. 
 References: 1 Dinescu+10, 2 Ortolani+11, 
3 Rossi+15a, 4 Dinescu+13, 5 Dinescu+03, 6 Zoccali+01 7 Cudworth\&Hanson93.
Notes: *: sample from Barbuy+98; ** papers from Ortolani+ along the years}
\small
\begin{center}
\begin{tabular}{l@{}r@{}r@{}r@{}r@{}r@{}r@{}r@{}r@{}r@{}r@{}r@{}r@{}r@{}r@{}r@{}r@{}r@{}r@{}}
\hline\hline
Name  & l($^{\circ}$) & b($^{\circ}$)
& \hbox{E(B-V)}  & E(B-V) & E(B-V)&  d$^{Harris}_{\odot}$ &d$^{Valenti}_{\odot}$& \hbox{d$_{\rm GC}$} & \phantom{-}\hbox{[Fe/H]}
& \hbox{v$_{r}$} & \hbox{v$_{s}$} & \\
  &  & 
& $^{Harris}$ & $^{Valenti}$ & $^{Ortolani}$ &  (kpc) & (kpc)& (kpc) & 
& \hbox{km s${-1}$} & \hbox{km s${-1}$} & \\
\hline%
 Terzan 3 & 345.08  &  9.19  & 0.73 & 0.73 & 0.72 & 8.20 & 8.10 &  2.5 &-0.72  &-136.3 &        &\\
 ESO452-SC11 & 351.91  &  12.10 & 0.46 & ---  & 0.58 & 8.30 & ---  & 2.1 &-1.50  &---    &          &\\
 NGC 6256 & 347.79  &   3.31 & 1.09 & 1.20 & 1.10 &10.30 & 9.10 & 3.0 &-1.02  &-101.4 &           &\\
 NGC 6266 (M 62)   & 353.58  &   7.32 & 0.47 & 0.47 & ---  &6.80  & 6.60 & 1.7  &-1.18  &-70.1  &           &\\
 NGC 6304 &  355.83 &   5.38 & 0.54 & 0.58 & 0.50 & 5.90 & 6.00 & 2.3 &-0.45  &-107.3 &195.69$^5$ &\\
 NGC 6316 &  357.18 &   5.76 & 0.54 & 0.56 & ---  &10.40 &11.60 & 2.6 &-0.45  & 71.4  &128.38$^5$ &&\\
 NGC 6325 &  0.97   &   8.00 & 0.91 & ---  & 0.95 & 7.80 & ---  & 1.1 &-1.25  &  29.8 &           &\\
 NGC 6342 &  4.90   &   9.73 & 0.46 & 0.57 & ---  & 8.50 & 8.40 & 1.7 &-0.55  &115.7  &155.26$^1$ &\\
 NGC 6355 &  359.58 &   5.43 & 0.77 & 0.81 & 0.78 & 9.20 & 9.00 & 1.4 &-1.37  &-176.9 &           &\\
 Terzan 2* (HP 3)   &  356.32 &   2.30 & 1.87 & 1.87 & 1.54 & 7.50 & 7.40 & 0.8 &-0.69  & 109.0 &141.87$^3$ &\\
 Terzan 4* (HP 4)   &  356.02 &   1.31 & 2.00 & 2.05 & 2.35 & 7.20 & 6.70 & 1.0 &-1.41  &-50.0  &172.17$^3$ &\\
 HP 1* (BH 229) &  357.42 &   2.12 & 1.12 & 1.18 & 1.19 & 8.20 & 6.80 & 0.5 &-1.00  &45.8   &241.45$^2$ &\\
 Liller 1* &  354.84 &  -0.16 & 3.07 & 3.09 & 3.05 & 8.20 & 7.90 & 0.8  &-0.33  & 52.0  &           &\\
 Terzan 1* (HP 2)   &  357.57 &   1.00 & 1.99 & 1.99 & 1.67 & 6.70 & 6.60 & 1.3 &-1.03  & 114.0 &125.36$^3$ &\\
 Ton 2 (Pis 26)    &  350.80 &  -3.42 & 1.24 & ---  & 1.26 & 8.20 & ---  & 1.4 &-0.70  &-184.4 &           &\\
 NGC 6401 &  3.45   &   3.98 & 0.72 & 1.10 & 0.53 &10.60 & 7.70 & 2.7 &-1.02  & -65.0 &           &\\
 VVV-CL002 & 359.56  & 0.89   & 2.88 & ---  & ---  & 7.3  & ---  & 0.7        &-0.40  &---    & &	   &\\
 Pal 6* &  2.09   &   1.78 & 1.46 & ---  & 1.33 & 5.80 & ---  & 2.2 &-0.91  &181.0  &219.67$^3$ &\\
 Djorg 1* &356.67 & -2.48    &  --- & ---  & ---  & ---  & ---  &        &--- 
 &---    & &	   &\\
 Terzan 5* &  3.81   &   1.67 & 2.28 & 2.38 & 2.49 & 6.90 & 5.90 & 1.2 &-0.23  &-93.0  &           &\\
 NGC 6440 &  7.73   &   3.80 & 1.07 & 1.15 & 1.00 & 8.50 & 8.20 & 1.3 &-0.36  &-76.6  &           &\\
 Terzan 6* (HP 5)   &  358.57 &  -2.16 & 2.35 & 2.35 & 2.24 & 6.80 & 6.70 & 1.3 &-0.56  & 126.0 &           &\\
 Terzan 9* &  3.60   &  -1.99 & 1.76 & 1.79 & 1.95 & 7.10 & 5.60 & 1.1 &-1.05  & 59.0  &126.35$^3$ &\\  
 ESO456-SC38* (Djorg 2)  &  2.76   &  -2.50 & 0.94 & 0.94 &
 0.89 & 6.30 & 7.00 & 1.8 &-0.65  &---    &           &\\
 Terzan 10* &  4.42   &  -1.86 & 2.40 & ---  & 2.48 & 5.80 & ---  & 2.3 &-1.00  & ---   &           &\\
 NGC 6522*  &  1.02   &  -3.93 & 0.48 & 0.66 & 0.55 & 7.70 & 7.40 & 0.6 &-1.34  &-21.1  &\phantom{-}140.61$^3$, 235.60$^5$ &\\
 NGC 6528* & 1.14   &  -4.17 & 0.54 & 0.62 & 0.55 & 7.90 & 7.50 & 3.0 &-0.11  & 206.6 &198.75$^5$ &\\
 NGC 6539 & 20.80  &   6.78 & 1.02 & 1.08 & ---  & 7.80 & 8.40 & 3.0 &-0.63  & 31.0  &           &\\
 NGC 6540* (Djorg 3) &  3.29   &  -3.31 & 0.66 & 0.66 & 0.60 & 5.30 & 5.20 & 2.8 &-1.35  &-17.7  &157.83$^3$ &\\
 NGC 6553* &  5.25   &  -3.02 & 0.63 & 0.84 & 0.63 & 6.00 & 4.90 & 2.2 &-0.18  & -3.2  &225.61$^5$,230$^6$    &\\
 NGC 6558 & 0.20   &  -6.03 & 0.44 & ---  & 0.38 & 7.40 & ---  & 1.0 &-1.32  &-197.2 &188.66$^3$ &\\
Kronberger~49  &7.63 & -2.01    &  --- & --- & 1.35 & ---  & ---& 1.19 & -0.10
 & ---    & --- & \\
  AL 3* (BH 261)  &  3.35   &  -5.26 & 0.36 & ---  & 0.36 & 6.50 & ---  & 1.7     &---    &  ---  &           &\\
 GLIMPSE02 &  14.13  &  -0.64 & 7.85 & ---  & ---  & 5.50 & ---  & 3.0 &-0.33  & ---   &           &\\
Mercer 5  &17.59 & -0.86    &  --- & ---  & ---  & ---  & ---  & 2.45 & -0.86
 & ---    & --- & \\
 NGC 6624 &  2.79   &  -7.91 & 0.28 & 0.28 & ---  & 7.90 & 8.40 & 1.2 &-0.44  & 53.9  &           &\\
 NGC 6626 (M 28)   &  7.80   &  -5.58 & 0.40 & ---  & ---  & 5.50 & ---  & 2.7 &-1.32  & 17.0  &131.26$^4$,75.07$^7$ &\\
 NGC 6638 & 7.90   &  -7.15 & 0.41 & 0.43 & ---  & 9.40 &10.30 & 2.2 &-0.95  & 18.1  &           &\\
 NGC 6637 (M 69)   &  1.72   & -10.27 & 0.18 & 0.14 & ---  & 8.80 & 9.40 & 1.7 &-0.64  & 39.9  &           &  \\
 NGC 6642 &  9.81   &  -6.44 & 0.40 & 0.60 & 0.42 & 8.10 & 8.60 & 1.7 &-1.26  & -57.2 &           &\\
 NGC 6652 &  1.53   & -11.38 & 0.09 & ---  & 0.10 &10.00 & ---  & 2.7 &-0.81  &-111.7 &           &\\
 NGC 6717 (Pal 9)  &  12.88  & -10.90 & 0.22 & ---  & 0.23 & 7.10 & ---  & 2.4 &-1.26  & 22.8  &           & \\
 NGC 6723 & 0.07   & -17.30 & 0.05 & ---  & ---  & 0.87 & ---  & 2.6 &-1.10  & -94.5 &171.11$^5$ &\\
\hline\hline
\end{tabular}
\end{center}
\label{tab1}
\end{table*}

\section{Kinematics and orbits}

According to Minniti \& Zoccali (2008), the bulge kinematics as viewed from
the field stars lies between a purely rotational system and a velocity dispersion
supported one. Another important feature of stellar kinematics is the presence of
 a massive bar in the bulge (Blitz \& Spergel 1991). An X-shape of the bulge
 related to the box/peanut configuration has been suggested by McWilliam \& Zoccali (2010)
and Nataf et al. (2010), 
 from the double red clump detected in the near-IR CMDs of the bulge fields at Galactic
 latitudes $\|$b$\|$ $>$ 5.5$^{\circ}$. This is interpreted as secular evolution of bars (e.g. Athanassoula 2005)
 and leads to the idea that the galactic bulge is not a “classical” bulge.
 This issue is developed in other reviews in this volume.
 Observations of the X-shape profile 
are detailed thoroughly in Wegg et al. (2015 and references therein). 
It is a generic feature of boxy bulges 
and is seen in many other galaxies (e.g. Bureau 2006).

Even if the X-shape is a confirmed feature, it is interesting to point out
a  recent study by Lee et al. (2015), that suggests an alternative explanation based on the
 presence of two different populations with a second generation of stars helium-enhanced and
 more metal rich, having a brighter clump than the first, more metal poor component. 
In this model there is no need for a deviation from a classical bulge shape. This is basically
 the same framework currently accepted for the multi-population features in the massive galactic 
globular clusters. Some issues are still open in this new interpretation, such as the source
 and the efficiency of the enrichment mechanism and the needed yields for the second generation component, 
in a wide environment such as the Galactic bulge. Accurate kinematics of the two clumps 
(for example from GAIA) could disentangle between the two scenarios.

\subsection{Bulge field properties}

{\it Kinematics.} Babusiaux et al. (2010, 2014) carried out a kinematical study of 650
 bulge field stars, and concluded that the more metal-poor stars,
where a lower end at around [Fe/H]$\approx$-1.0 was found,
correspond to a spheroidal distribution. The metal-rich stars showed instead
a kinematics typical of the bar (see also review by Gonzalez \& Gadotti 2015). 
 V\'asquez et al. (2013) studied the kinematics of 454 field bulge stars located
in the bright and the faint red clumps of the X-shaped bulge. They conclude that
stars with elongated orbits tend to be metal-poor, whereas the metal-rich ones are
preferentially in axisymmetric orbits, at odds
 with conclusions by Babusiaux et al. (2010).

{\it Ages.} 
Clarkson et al. (2008, 2011) used proper motion cleaned
data, deriving a cleaned bulge CMD, demonstrated to consist of an old population of
at least 10 Gyr. The same conclusion had been reached previously by Zoccali et al. (2003).

 RR Lyrae trace an old spheroidal component (Lee et al. 1992; D\'ek\'any et al. 2013;
Pietrukowicz et al. 2012, 2015), with a peak in metallicity at [Fe/H]$\sim$-1.0. This old
stellar population is compatible with the metal-poor spheroidal component studied by
Babusiaux (2010, 2014). 
The moderately metal-poor clusters like NGC 6522 should belong to a same
population as these RR Lyrae.

\subsection{Kinematics and orbits of bulge GCs}

Space velocities for the GCs, that require radial velocities and proper motions to be derived,
are only available for part of the sample.
Tables  \ref{tab1} and \ref{tab2} gather space velocities with respect to the Sun,
 and corresponding references. These space velocities, which result from a combination
of radial velocities and proper motions, are reported in column 12 of \ref{tab1} and column 8 of
\ref{tab2}.
 Earlier work was carried out by
Dinescu et al. (1997, 1999a,b, 2003) and Casetti-Dinescu et al.
(2007, 2010, 2013).
Using Hubble Space Telescope data,  Zoccali et al. (2003) and Feltzing
\& Johnson (2002) derived space velocities for the metal-rich clusters NGC 6553 and
NGC 6528 respectively. Rossi et al. (2015a) measured proper motion
cleaned CMDs from long time base data, and 
derived  space velocities for 10 central globular clusters. 

An important piece of information was revealed by the orbits of the inner
GCs derived by Rossi et al. (2015b): all GCs located in the inner bulge appear 
to be trapped in the bar (Rossi et al. 2015a,b). We point out that the clusters, in particular the moderately
metal-poor ones, probably formed very early in the very central parts of the Galaxy
(e.g. NGC 6522, NGC 6558), before the bar instability occurred. This is confirmed by their
rotational velocity counter-rotating with respect to the bar and the Galaxy (Rossi et al.
 2015b). Since these clusters have low heights z, their retrograde
orbits can be considered as robust with respect to variations of the bar shape
(Pfenniger 1984).
Therefore, it seems that whenever the bar formed, given their low kinematics,
 essentially all GCs would be trapped.
 Martinez-Valpuesta (2006)
 suggests that a first vertical buckling of the growing bar occurred at 
1.8-2.8 Gyr, and a second at 6-7.5 Gyr. The bar then assumes a boxy
or peanut X-shape.  The trapping of GCs in the boxy bulge includes
bulge clusters of all metallicities, from
moderately metal-poor as mentioned above,
to metal-rich ones like Terzan 2 (Rossi et al. 2015a,b).
 
A main conclusion is that the inner clusters are confined and possibly 
trapped in the bar, due to the high mass of the bar, achievable due to the low
velocity components of the clusters. Orbits computed by Rossi et al. (2015b)
showed that the inner clusters have  confined orbits as compared with the 
outer bulge/inner halo shell.

\section{Future steps\label{sect7}}

Figure \ref{Fig5} shows the globular clusters projected in the
central $\|$l$\|$$<$8$^{\circ}$ and $\|$b$\|$$<$5$^{\circ}$. These include 
the sample by  Barbuy et al. (1998), 
which was given in a radius of R$<$5$^{\circ}$,
  and a few more, in particular
Terzan 9, Al~3, and results from the VVV survey. The  
VVV survey provided  new GCs and candidates in that region
 (Borissova et al. 2011, 2014). 
The clusters VVV CL001, CL 002, CL 003 and  CL 004
were studied by means of the VVV photometric catalogue  (Minniti et  
al. 2011, Moni Bidin et al.
2011). VVV CL 002 appears to be the most central GC. VVV CL 001 is  
projected very close to UKS 1 and they may be a
binary system (Minniti et al. 2011). In this case
 UKS~1 might be interpreted as  a  
dwarf galaxy remains, similarly to Terzan 5.
 VVV CL 003 seems to be a far side GC or old  
open cluster, while VVV CL 004  might be rather an old open cluster.
Table \ref{tab4} shows  available information on the VVV sample, including ages
and  GC candidates (Borissova et al. 2014).
Table \ref{tab4} reveals a number of candidates in zones where 2MASS
and GLIMPSE could not detect clusters. These previous surveys showed
GCs outside the central bulge only.

Fig. \ref{Fig5}  shows the angular distribution of known GCs 
projected in the central parts of the Galaxy, and candidate ones found
in the VVV survey. This figure, together with Table \ref{tab1},
 clearly show  the depletion of GCs on the far side.  
Besides,  VVV CL 002 is very close to the Galactic
center, and VVV CL 003 is in the far size. The VVV  
candidates (Table \ref{tab4}) may  mitigate that
asymmetric distribution, but will not completely solve the 
problem of missing globular clusters in the far side.
We recall that however when IR distances are taken into account,
the clusters are well distributed around 8 kpc (Sect. \ref{sect3}).

Ivanov, Kurtev \& Borissova (2005) estimated that in the central parts  
of the Galaxy  at least 10$\pm$3 GCs were missing.
Several new GCs have been  discovered or identified as such in the  
last decade, especially with observations in the near IR, in the area
outside Fig. \ref{Fig5}, and recently, in the inner bulge region  VVV CL 002 has  
been added. The  VVV candidates are a promising sample
to further populate the inner bulge.

Fig. \ref{Fig5}  portrays  the current status of the  inner bulge sample,
together with  VVV GCs and candidates (Table \ref{tab1}). As pointed out by  
Barbuy et al. (1998)
a zone of  avoidance in the GC distribution  occurs for 0.9 $<$ b $<$  
-1.9. It is  related to dust
heavily  absorbing in the disk and/or dynamical effects on the GC  
population by the disk (bar) and bulge.
The zone of avoidance  is asymmetric in Galactic latitude  because of  
the Sun's offset of about 18 pc above
the Galactic plane. The zone of avoidance now includes 5 VVV  
candidates (\ref{Fig5}), while the other  ones are
distributed as in the sample available previously in Barbuy et al. (1998).

Finally, as concerns a possible binarity between VVV CL 001 and UKS~1,
it is interesting to note that, 
likewise, the halo dwarf spheroidal Ursa Minor has a GC companion
(Mu\~noz et al. 2012).
Growing evidence suggests that a  fraction of the
inner bulge sample are galaxy remains. In the coming years
GAIA may show streamers related to them, to further constrain
the dynamical issues involved.

\begin{table*}
\small
\caption{Intruders and missed bulge clusters?
a) Possible halo intruders with [Fe/H]$<$-1.5; 
b) Shell: Distances 3$<$R$<$4.5 kpc;
c) Intruders to shell; d) [Fe/H]$>$-1.0 and R$>$4.5kpc;
e) Intruders to disc; f) no parameters enough;
 * looks halo intruder in the shell,
 despite distance; ** see Sect. \ref{sect3}.
References: 1 Dinescu+10, 4 Dinescu+13 5 Dinescu+07, 
6 Dinescu99a, 7 Cudworth 93, 8 Dinescu+97, 9 Pe\~naloza+15
}
\begin{center}
\begin{tabular}{ccrrrrrrrrrrrr}
\hline\hline
Name & Name  & l($^{\circ}$) & b($^{\circ}$)
& \hbox{d$_{\rm GC}$ (kpc)} & \hbox{[Fe/H]}
& \hbox{v$_{r}$ km s$^{-1}$} & \hbox{v$_{s}$  km s$^{-1}$} \\
\hline%
\noalign{a) Probable halo intruders with [Fe/H]$<$-1.5}
\hline
NGC 6144 &  & 351.93 &  15.70 &2.7 &-1.76&  201.9 &366.70$^6$ &	   \\
NGC 6273 & M 19  & 356.87 &   9.38 &1.7 & -1.73 &143.8  &213.73$^1$ & \\
NGC 6287 &  & 0.13 & 11.02 &2.1 &-2.10& $-$279.1 &280.19$^1$ &	   \\
NGC 6293 &  & 357.62 & 7.83 &2.3 &-1.99&$-$137.4 &201.35$^1$ &	   \\
NGC 6333 & M9 & 5.54  & 10.70 &1.7 &-1.77 &239.8 &284.63$^1$	   \\
NGC 6541 &  & 349.29 & -11.18 & 2.1 &-1.81& $-$154.2 & &	   \\
\hline
\noalign {b) Outer Shell of distances 3$<$R$<$4.5 kpc}
\hline
Lynga 7 &    & 328.77  & -2.79  & 4.3 &-1.01 & 8.0 & &	   \\
NGC 6171 & M107 & 3.37  & 23.01  & 3.3 &-1.02 & $-$23.0 &87.01$^7$ &	   \\
NGC 6235 &  & 358.92 & 13.52 & 4.2 &-1.28 & 96.8 & &	   \\
NGC 6402 & M14 & 21.32  & 14.81 & 4.0 &-1.28 & $-$42.1 & &	   \\
NGC 6388 &  & 345.56 & -6.74 & 3.1 &-0.55 & 84.2 &43.30$^1$ &	   \\
NGC 6352 &  & 341.42  & -7.17 & 3.3 &-0.64 & $-$134.1 & &	   \\
NGC 6380 &Ton 1  & 350.18  & -3.42 & 3.3 &-0.75 & 2.1 & &	   \\
NGC 6441 &  & 353.53 & -5.01 & 3.9 &-0.46 & 22.9 &48.80$^1$ &	   \\
NGC 6496 &  & 348.02 & -10.01  & 4.2 &-0.46 & $-$108.4 & &	   \\
NGC 6517 &  & 19.23 & 6.76 & 4.2 &-1.23 &$-$26.5 & &	   \\
NGC 6544*& & 5.84 & -2.20  & 5.1 & -1.4  & $-$17.7 & &  \\
2MASS-GC02 &  & 9.78 & -1.08 & 3.9$^{8}$ &-1.08 &$-$227.4 & &	   \\
IC 1276 &Pal 7  & 21.83 & 5.67 & 3.7 &-0.75 & 169.2 & & \\
Terzan 12 &  & 8.36 & -2.10 & 3.4 &-0.50 & 104.2 & &	   \\
NGC 6569 &  & 0.48  & -6.68 & 3.1 &-0.76 & $-$20.3 & & \\
NGC 6712 &  & 25.35  & -4.32 & 3.5 &-1.02 & $-$94.7  &166.55$^7$ &	   \\
\hline
\noalign {c) Outer  shell intruders}
\hline
NGC 6139 &  & 342.37  & 6.94  & 3.6 &-1.65 & 11.5 & &	   \\
NGC 6453 &  & 355.72 & -3.87 &  3.7 &-1.50& $-$76.7 & &	   \\
NGC 6535 &  & 27.18 & 10.44 & 3.9 &-1.79 & $-$200.4 & &	   \\
NGC 6656 &M22 & 9.89  & -7.55  & 4.9 &-1.70 & $-$136.6 &277.62$^4$,187.27$^7$ &	   \\
NGC 6809 & M55 & 8.80  & -23.27  &  3.9 &-1.94& 181.8 & &	   \\
\hline
\noalign {d)  Metal-rich clusters ([Fe/H]$>$-1.0) beyond R$>$4.5kpc}
\hline
NGC 104   & 47 Tuc   & 305.90 & -44.89 & 7.4 &-0.72 & $-$26.7 &77.76$^7$ & \\
NGC 5927 &    & 326.60 & 4.86 & 4.6 &-0.49 & $-$107.2 &230.2$^5$ & \\
BH 176 &    & 328.41 & 4.34 & 12.9 &0.00 & --- & &	   \\
NGC 6356 &    & 6.72 & 10.22 & 7.5 &-0.40 &  38.0 &144.54$^1$ &	   \\
NGC 6362 &    & 325.55 & -17.57 & 5.1 &-0.99 & $-$15.6 &144.55$^8$ & \\
NGC 6366 &    & 18.41 & 16.04 & 5.0 &-0.59 & $-$108.6  & & \\
UKS 1*  & & 5.12   &   0.76  & 6.7$^{**}$  &-0.64  & 57.0 &           &\\
Pfleiderer 2 &  & 22.28 & 9.32 & 9.7 &0.0 & --- & &	   \\
Pal 8 &    & 14.10 & -6.80 & 5.5 &-0.37 & $-$32.3 & &	   \\
NGC 6760 &  & 36.11 & -3.92 & 4.8 &-0.40 & $-$13.2 & &	   \\
Pal 11 &    & 31.81 & -15.58 & 6.7 &-0.40 & $-$56.0 & &	   \\
NGC 6838 & M71 & 56.74  & -4.56 & 6.7 &-0.78 & $-$7.5 &96.88$^7$ &  \\
\hline
\noalign {e) intruders to d) }
\hline
NGC 6284 &  & 358.35 &  9.94 &7.5  & -1.26 &36.7&249.56$^1$ &  \\
FSR 1767 &  & 352.60 & -2.17 &6.53  & -1.20 &---&--- &  \\
\hline
\noalign{f) Little studied clusters without enough parameters}  
\hline
FSR 1735 & & 339.20 & -1.85 & 3.7 &--- &... & &	   \\
VVV-CL003 & & 358.40  & 0.73  &13.0 &-0.10 &... & &	   \\
VVV-CL001 & & 5.27  & 0.78  & --- &--- &... & &	   \\
VVV-CL004 & & 6.79  & 1.72  & --- &--- &... & &	   \\
2MASS-GC01 &  & 10.47 & 0.10 & 4.5 &---&...& &	   \\
GLIMPSE-C01 &  & 31.30 & -0.10 & 4.9 &--- &... & &	   \\
\hline\hline
\end{tabular}
\end{center}
\label{tab2}
\end{table*}

\begin{table*}
\small
\caption{Metallicities and abundances from high resolution spectroscopy. References:
  11 Origlia+05a; 
12 Origlia \& Rich04; 13 Barbuy+06,+15; 14 Origlia+02; 15 Valenti+15; 
16 Lee+04; 17 Origlia+11;
18 Origlia+08; 19 Origlia05b; 
20 Barbuy+14; 21 Zoccali+04; 22 Carretta+01;  23 Mel\'endez+03 plus Origlia+02; 24 Cohen+99;
25 Alves-Brito+06;
 26 Barbuy+07; 27 Valenti+11;
 28 Smith \& Wehlau85; 29 Lee+07; 30 Gratton+15 for BHB; 31 Gratton+15 for RHB; 32 Pe\~naloza+15. }
\begin{center}
\begin{tabular}{c@{}c@{}r@{}r@{}r@{}r@{}r@{}r@{}r@{}r@{}r@{}r@{}r@{}r@{}r@{}r@{}r@{}r@{}r@{}r@{}}
\hline\hline
Name  & \hbox{[Fe/H]}
& \hbox{[C/Fe} & \hbox{[N/Fe]} & \hbox{[O/Fe]} & \hbox{[Na/Fe]} & \hbox{[Al/Fe]} & \hbox{[Mg/Fe]} & \hbox{[Si/Fe]} & \hbox{[Ca/Fe]} & \hbox{[Ti/Fe]}  
& \hbox{[Y/Fe]} & \hbox{[Ba/Fe]} & \hbox{[La/Fe]} & \hbox{[Eu/Fe]} &\phantom{-}ref. & \\
\hline%
 NGC 6342 &-0.60 &-0.34 &-- &+0.31 &-- &-- &+0.38  & +0.37 &+0.38 &+0.25 &-- &-- &--  &11 &  \\
 Terzan 4 & -1.60   &-0.25 &-- &+0.54 &-- &-- &+0.41 &+0.55 &+0.54 &+0.44 &-- &-- &-- &-- & 12 &  \\
 HP 1     & -1.00   &-- &-- &+0.40 &+0.00 &-- &+0.10 &+0.30 &+0.03 &+0.02 &-- &+0.15
 &+0.00 &+0.15  & 13 & \\
 Liller 1 & -0.30   &-- &-- &-- &-- &-- &+0.3 &+0.3 &+0.3 &-- &--  &-- &-- &-- & 14 & \\
 Terzan 1 & -1.26   &-- &-- &+0.39 &-- &-- &+0.42 &+0.31 &-- &+0.15 &-- &-- &-- &-- & 15 &  \\
 Pal 6    & -1.00   &-- &-- &-- &-- &-- &-- &-- &-- &-- &-- &-- &-- &--  & 16&  \\
 Terzan 5 &-0.25 &-0.34 &-- &+0.28 &-- &-- &+0.30 &+0.35 &+0.31 &+0.31 &-- &-- &-- &-- &12 &  \\
Terzan 5 & +0.25  &-0.40 &-- &-0.05 &-- &-- &+0.10 &-0.02 &+0.00 &+0.05 &-- &--
 &-- &-- &17&  \\
 NGC 6440 & -0.50 &-0.57 &+0.33 &-- &+0.46 &+0.33 &+0.32 &+0.37 &+0.33 &-- &-- &-- &-- &-- & 18 & \\
 UKS 1    & -0.78   &-0.45 &-- &+0.28 &-- &-- &+0.32 &+0.28 &+0.38 &+0.32 &-- &-- &-- &-- &19 &  \\
 NGC 6522 & -0.95   &-0.03 &+0.67 &+0.36 &-0.07 &-0.11 &+0.23 &+0.13 &+0.13 &+0.10 &+0.31 &+0.02 &-0.01 &-0.14 & 20 & \\
 NGC 6528 &-0.17 &-0.11 &-- -- &+0.15 &+0.43 &-- &+0.07 &+0.08 &$-$0.40 &$-$0.10 &-- &-- &-- &-- &21&  \\
 NGC 6528 &+0.07 &-- &-- &+0.07 &+0.40 &-- &+0.14 &+0.36 &+0.23 &+0.03 &-- &+0.14 &-- &-- &22&  \\
 NGC 6528 &-0.17 &-0.35 &-- &+0.33 &-- &-- &+0.35 &+0.30 &+0.37 &+0.31 &-- &-- &-- &-- &11&  \\
 NGC 6539 & -0.76      &-0.30 &-- &+0.41 &-- &-- &+0.45 &+0.43 &+0.43 &+0.42 &--
 &-- &-- &-- &19 &  \\
 NGC 6553 & -0.20$^4$   & -0.08$^4$&+0.30$^4$ &+0.43$^4$ &-- &-- &+0.3$^5$ &+0.3$^5$ &+0.3$^5$ &-- &-- &-- &-- &-- &23&  \\
 NGC 6553 & -0.17   & --&-- &+0.50 &-- &-- &+0.41 &+0.14 &+0.26 &+0.19 &-- &-- &-- &-- &24&  \\
 NGC 6553 & -0.21$^4$   & --&-- &-- &+0.16 &+0.18 &+0.28 &+0.21 &+0.05 &$-$0.01 &-- &$-$0.28
 &$-$0.11 &+0.10 &25&  \\
 NGC 6558 & -0.97   &-- &-- &+0.38 &-0.09 &+0.02 &+0.24 &+0.23 &+0.05 &+0.06 &-- &+0.13 &0.00 &+0.36 & 26 & \\
 Mercer 5 & $-$0.85 &--- &--- &+0.31 &---   & ---& ---  &+0.53 &---   &+0.42 &---&---   &---&---&32 & \\
 NGC 6624 & -0.69  &-0.29 &-- &+0.41 &-- &+0.39 &+0.42 &+0.38 &+0.40 &+0.37 &-- &-- &-- &-- &27 &  \\
 NGC 6626 &-1.00 &--&-- &-- &-- &-- &-- &-- &-- &-- &-- &-- &-- &--  &28&  \\
 NGC 6637 &$-$0.77 &--- &--- &+0.20 &+0.35 &+0.49
 &+0.28 &+0.45 &+0.20 &+0.24 &+0.13 &+0.22 &+0.21 &+0.45 &29 &  \\
 NGC 6723 & $-$1.22 &--- &+0.85 &+0.39 &+0.04 &--- &+0.52 &-- &-- &-- &-- &--  &-- &-- & 30 & \\
 NGC 6723 & $-$1.22 &--- &--- &+0.55 &+0.11 &--- &+0.50 &+0.59 &+0.81 &--    &-- &+0.75 &-- &-- & 31 & \\
\hline\hline
\end{tabular}
\end{center}
\label{tab3}
\end{table*}

\begin{table*}
\caption{ VVV GCs and candidates. References:  
(1) Minniti et al.(2012), (2) Moni Bidin et al. (2011), (3)  
Borissova et al. (2014).
}
\begin{center}
\begin{tabular}{ccrrrrrrrrrrrr}
\hline\hline
VVV    &    l  &   b &   alpha  &   delta &  Appr.size &   comments &  References & \\
       &    (o) &  (o) &  h:m:s &     (o):':'' &  (')  &  & \\
  1    &    2   &  3   &    4   &      5     &  6    & 7 & 8 & \\           
\hline
CL110  &  352.54 &  0.95 &  17:22:47 & -34:41:17 &  0.4  &   GC/old OC & 3 & \\ 
CL119  & 355.16  & 0.71 & 17:30:46 & -32:39:05 & 1.8   & old OC     & 2  & \\   
CL003  &  358.40 & 0.73 & 17:38:55 & -29:54:25 &  2.2  &   old OC/GC,far side(13 kpc) & 2 & \\ 
CL128  &  356.38 & -0.81 & 17:39:59 & -32:26:27 &  1.0  &   GC cand./Old OC & 3 & \\     
CL002  & 359.56  & 0.89 & 17:41:06 & -28:50:42 &  3.0   &      GC   & 2 & \\           
CL131  &  354.72 & -2.17 & 17:41:17 & -34:34:02 &  1.7  &   old OC/GC cand.  & 3 & \\    
CL143  & 355.79  & -2.32 & 17:44:36 & -33:44:18 & 1.3  & old OC/young GC & 2 & \\
CL150  &   3.77 & 0.96 & 17:50:41 & -25:13:06  & 0.8  &   old OC/young GC,10Gyr & 3& \\
CL154  &   1.79 & -1.37 & 17:55:08 & -28:06:01 &  0.7  &   old OC, 8 Gyr & 3 & \\
CL004  &   6.79 & 1.72 & 17:54:32 & -22:13:38 &  1.6  &   old OC?  & 2 & \\ 
CL001  &   5.27 & 0.78 & 17:54:43 & -24:00:53 &  0.5  &   GC,8´ from UKS 1 &  1 & \\   
\hline\hline
\end{tabular}
\end{center}
\label{tab4}
\end{table*}

\section{Conclusions}

In this review we provide a state-of-the-art list of bulge globular
clusters and their properties. 
We tried to constrain their properties including their kinematics
when available. Kinematics is becoming a key tool to identify their nature,
and the link with the field stellar populations, within the complex
substructures of the bulge. 

In recent years  good progress in the knowledge of the globular clusters
  in the
central parts of the Galaxy has been achieved. For spectroscopy the multi-object
spectrographs
in 8m-class telescopes have made it possible to derive chemical abundances
for considerable numbers of stars. Progress in instrumentation thanks to e.g.
 imaging with MCAO in the infrared
(MAD/VLT, GEMS/Gemini) and HST/NICMOS, made possible  long time baseline of
CCD data with excellent quality, allowing proper motion cleaning of CMDs.
Finally, surveys with larger
apertures such as the 4m VISTA Telescope, provided the discovery of new objects. 
The VVV survey in particular
has provided several new bulge globular clusters and candidates, to
be explored in coming years.

 Much work is still needed as concerns bulge globular clusters, such as
the monitoring of variable stars, in particular RR Lyrae, 
 to obtain deep CMDs allowing age derivation, and the derivation
of metallicities and chemical abundances from high-resolution spectroscopy.
The derivation of Na and O abundances will be crucial to define
if these very old clusters have a unique or multiple stellar generations.
These results should allow to better compare the bulge globular clusters
 with outer bulge, inner halo and outer halo ones.


\begin{acknowledgements}
EB and BB acknowledge partial financial support from CNPq, CAPES and Fapesp.
SO acknowledges the Italian Ministero dell'Universit\`a e della Ricerca
Scientifica e Tecnologica (MURST), Italy. 
\end{acknowledgements}



\end{document}